Quantized representation for Kadomtsev-Petviashvili equation on the soliton sector

Yair Zarmi

Jacob Blaustein Institutes for Desert Research
Ben-Gurion University of the Negev
Midreshet Ben-Gurion, 84990 Israel

Abstract

Exploiting the known structure of soliton solutions, obtained through the Hirota transformation, a quantized representation of the Kadomtsev-Petviashvili (KP) equation on the soliton sector is constructed over a Fock space of particles, which may be either bosons or fermions. The classical solution is mapped into an operator, which also obeys the KP equation. The operator is constructed in terms of the particle-number operators. Classical soliton solutions are the expectation values of this operator in multi-particle states in the Fock space. The operator is equation-specific and the state in the Fock space is in one-to-one correspondence with the particular soliton solution.

# 1. Introduction

The quantization of classical integrable non-linear evolution equations, and, more generally, the role of solitons in quantum field theories have attracted much attention in recent decades [1-63]. Employing concepts and powerful tools, borrowed from the formalisms of Inverse Scattering, Field Theory, Symmetry Groups and S-Matrix Theory, standard canonical quantization procedures have led to formal quantization of quite a few equations. Prominent amongst them in (1+1) dimensions are the Korteweg-de Vries [6-10, 12, 13, 16-19, 21, 22, 25, 29, 30, 33, 40-46, 48, 49, 51-53, 56], Non-linear Schrödinger [1, 16, 42, 50, 55, 63], Sine-Gordon [16, 34, 57, 61] and Nonlinear Klein-Gordon [23] equations, and, in (2+1) dimensions, the Kadomtsev-Petviashvili equation [2-6, 68, 18, 20, 27, 28, 31, 32, 38, 45].

Traditionally, integrable non-linear evolution equations have been derived as approximate descriptions of particular physical systems, and their soliton solutions were then obtained through either direct computation, or, most often, with the aid of the formalism of the Inverse Scattering Transform. A fundamental change has occurred with the development of an approach combining algebraic and field-theoretical methods [2-6, 9, 12, 13, 14, 16, 18-21, 23, 27, 28, 30, 32, 33, 38, 40-42, 51-54]. Many of the known integrable non-linear evolution equations that have soliton solutions emerge as group-theoretic constructs related to the equations that define infinite-dimensional Grassman manifolds. In addition, the classical soliton solutions naturally emerge in terms of group-theoretical entities [2-6], e.g., as the vacuum expectation values of operators

$$u = \langle 0 | \mathrm{J} | 0 \rangle \ , \tag{1}$$

or as matrix elements of operators in the highest weight vectors of representations of a Lie algebra

$$u = \langle l | \mathrm{K} | l \rangle \ , \tag{2}$$

or, equivalently, as linear combinations of characters of the $GL(N)$ groups.

Eqs. (1) and (2) imply that for each evolution equation, and for each soliton solution, one has to construct a different operator, J or K.

Recently, an algorithm for the construction of a simple quantized representation over the soliton sector for integrable non-linear evolution equations in (1+1) dimensions, which can be solved through the Hirota transformation [64], has been proposed in [89]. Exploiting the known structure of classical $N$-soliton solutions, it is shown that they can be obtained as matrix elements of an operator in an $N$-particle state in the Fock space:

$$u = \langle \psi_N | \mathrm{U} | \psi_N \rangle \quad . \tag{3}$$

The particles may be either bosons or fermions. The operator, U, is constructed in terms of the particle-number operators. It is characteristic of the equation, and solves it. Moreover, apart from equation-specific coefficients, the structure of the operator is the same for all the equations in (1+1) dimensions that are solved by the same Hirota transformation. In [89] this is demonstrated for the KdV [65], Sawada-Kotera [66, 67] and bi-directional KdV [68, 67] equations. The Hirota transformation that generates soliton solutions of the mKdV equation is different [69, 70]. Consequently, so is the operator, U. In all cases, the state, $|\psi_N\rangle$, uniquely determines the soliton solution. Finally, the algorithm allows for the inclusion of interactions that do not have a classical analog (e.g., annihilation and creation of solitons), hence, allowing for the study of nonlinear quantum-dynamical systems, for which the lowest-order approximation is known, and is associated with the classical soliton solutions. An example to this effect is presented in [89].

In this paper, the quantized representation of soliton solutions leading to Eq. (3) is constructed for the Kadomtsev–Petviashvili (KP) equation [71]:

$$\frac{\partial}{\partial x}\left(-4\frac{\partial u}{\partial t} + \frac{\partial^3 u}{\partial x^3} + 6u\frac{\partial u}{\partial x}\right) + 3\frac{\partial^2 u}{\partial y^2} = 0 \quad . \tag{4}$$

Eq. (4) approximately describes the propagation of disturbances on the surface of a shallow-water layer in terms solitons in (2 +1) dimensions. It is integrable, and its soliton solutions, $u(t,x)$ are constructed in terms of the Hirota transformation [72]:

$$u(t,x,y) = 2\frac{\partial^2}{\partial x^2}\ln f(t,x,y) \ . \tag{5}$$

Construction of the quantized representation depends on the structure and properties of the soliton solutions of Eq. (4). As these have been studied extensively in the literature, they are merely reviewed in Section 2. The construction of the quantized representation is presented in Section 3.

## 2. Construction and properties of soliton solutions – a review
### 2.1 Structure of soliton solutions

The extensive literature [72-86] on soliton solutions of Eq. (3) can be summarized as follows. First, one selects a set of $M$ different wave numbers:

$$q_1 < .... < q_M \ . \tag{6}$$

The function $f(t,x,y)$ in Eq. (5) is a sum of terms, each of which is a product of $N$ exponential factors ($N < M$). Based on the observations of [72-82], $f(t,x,y)$ may be written as:

$$f(t,x,y) = \begin{cases} \sum_{i=1}^{M}\xi_M(i)\exp(\theta_i) & N=1 \\ \sum_{1\le i_1<....<i_N \le M}\xi_M(i_1,.....,i_N)\left(\prod_{1\le j<l\le N}(q_{i_l}-q_{i_j})\right)\exp\left(\sum_{j=1}^{N}\theta_{i_j}\right) & 2\le N\le M-1 \end{cases}$$

$$\equiv f_{\vec{\xi}}(M,N) \tag{7}$$

The exponents are given by:

$$\theta_i(t,x,y) = -q_i x + q_i^2 y - q_i^3 t + \theta_i^0 \ . \tag{8}$$

Here $q_i$ are wave numbers, $\theta_i^0$ are arbitrary phase shifts, and the coefficients, $\xi_M(i_1,.....,i_N)$, are the components of a vector, denoted by $\vec{\xi}_M$. In each term in $f_{\vec{\xi}}(M,N)$, the $N$ wave numbers con-stitute one of the $\binom{M}{N}$ subsets of the pre-selected set, $\{q_1,...,q_M\}$. Finally, to exclude singular solutions of Eq. (4), one requires

$$\xi_M(i_1,.....,i_N) \geq 0 \ . \tag{9}$$

The notation $f_{\bar{\xi}}(M,N)$, introduced in Eq. (4), will be used in the following, and the corresponding solution of Eq. (3) will be denoted by $u_{\bar{\xi}}(M,N)$.

## 2.2 Space-time reflection invariance and structure of solutions
Invariance of Eq. (4) under space-time reflection, $(t,x,y) \to (-t, -x, -y)$, implies that if $u(t,x,y)$ is a solution of Eq. (1), so is $u(-t, -x, -y)$. This property is reflected in the structure of the soliton solutions. For $f_{\bar{\xi}}(M,N)$ of Eq. (7), which generates a solution, $u_{\bar{\xi}}(M,N)$, of Eq. (4), define

$$\tilde{f}_{\bar{\xi}}(M,N) = \frac{f_{\bar{\xi}}(M,N)}{\prod_{i=1}^{M} \exp\theta_i} \ . \tag{10}$$

Under space-time reflection, $\tilde{f}_{\bar{\xi}}(M,N)$ is transformed into

$$\tilde{f}_{\bar{\xi}}(M,N) \xrightarrow[(t,x,y)\to(-t,-x,-y)]{} f_{\bar{\xi}}(M, M-N) \ , \tag{11}$$

where $f_{\bar{\xi}}(M, M-N)$ generates a solution, $u_{\bar{\xi}}(M, M-N)$, of Eq. (4). Thus, all solutions with $[M/2] \leq N \leq M-1$ may be obtained from solutions with $1 \leq N \leq [M/2]$ by space-time reflection.

## 2.3 Unique nature of (M,1) and (M,M – 1) solutions
Apart from the requirement that they are non-negative, all coefficients in $f_{\bar{\xi}}(M,1)$ may obtain *arbitrary* values. Due to Eq. (11), all coefficients in $f_{\bar{\xi}}(M, M-1)$ may also obtain *arbitrary* values.

## 2.4 The case $M \geq 4$, $2 \leq N \leq M-2$
For $M \geq 4$ and $2 \leq N \leq M-2$, the coefficients, $\xi_M(i_1,.....,i_N)$ are given in terms of matrix elements of an $N \times M$ ($N \leq M$) matrix [72-82]:

$$A = (a_{ji}) \quad , \quad (1 \le j \le N, 1 \le i \le M) \quad . \tag{12}$$

$\xi_M(i_1,....,i_N)$ is the determinant of the $N \times N$ minor given by the $i_j$'th columns with $j = 1, \ldots, N$:

$$\xi(i_1,...,i_N) = \begin{vmatrix} a_{1,i_1} & \cdots & a_{1,i_N} \\ \vdots & \ddots & \vdots \\ a_{N,i_1} & \cdots & a_{N,i_N} \end{vmatrix} . \tag{13}$$

The matrix $A$ must be constructed so that Eq. (9) is obeyed.

Once the structure of the soliton solutions is known, one may start from Eqs. (5) and (7), and search for the constraints that the coefficients, $\xi_M(i_1,....,i_N)$, must obey in order to obtain a valid solution. Direct computation yields that the must obey a number of bilinear constraints (the Plücker relations, see. E.g. [87]). For example, for $(M,N) = (4,2)$ one finds

$$\xi_4(1,2)\xi_4(3,4) - \xi_4(1,3)\xi_4(2,4) + \xi_4(1,4)\xi_4(2,3) = 0 \quad . \tag{14}$$

For $(M,N) = (5,2)$, one finds the following constraints on the $\xi$ − coefficients:

$$\begin{aligned}
\xi_5(1,4)\xi_5(2,3) - \xi_5(1,3)\xi_5(2,4) + \xi_5(1,2)\xi_5(3,4) &= 0 \\
\xi_5(1,5)\xi_5(2,3) - \xi_5(1,3)\xi_5(2,5) + \xi_5(1,2)\xi_5(3,5) &= 0 \\
\xi_5(1,5)\xi_5(2,4) - \xi_5(1,4)\xi_5(2,5) + \xi_5(1,2)\xi_5(4,5) &= 0 \\
\xi_5(1,5)\xi_5(3,4) - \xi_5(1,4)\xi_5(3,5) + \xi_5(1,3)\xi_5(4,5) &= 0 \\
\xi_5(2,5)\xi_5(3,4) - \xi_5(2,4)\xi_5(3,5) + \xi_5(2,3)\xi_5(4,5) &= 0
\end{aligned} \tag{15}$$

Eq. (13) provides a solution of the bilinear constraints that the $\xi$ - coefficients must obey. For example, identifying $\xi_M(i,j)$ for $(M,N) = (4,2)$ with determinants of 2×2 minors of a 2×4 matrix, and for $(M,N) = (5,2)$ - with determinants of 2×2 minors of a 2×5 matrix, one finds that $\xi_M(i,j)$ obey Eqs. (14) and (15), respectively [85]. However, the constraints may have other solutions.

The fact that, the $\xi$ - coefficients must obey some constraints implies that they cannot be absorbed in the arbitrary initial phases, $\theta_i^0$, in Eq. (8). As an example, consider the $(M,N) = (4,2)$ case, for which $f_{\bar{\xi}}(4,2)$ is given by:

$$f_{\bar{\xi}}(4,2) = \xi_4(1,2)e^{\theta_1+\theta_2} + \xi_4(1,3)e^{\theta_1+\theta_3} + \xi_4(1,4)e^{\theta_1+\theta_4} + \xi_4(2,3)e^{\theta_2+\theta_3} + \xi_4(2,4)e^{\theta_2+\theta_4} + \xi_4(3,4)e^{\theta_3+\theta_4} \quad (16)$$

Let us choose, for example,

$$\xi_4(1,2) = \xi_4(3,4) = \xi_4(1,4) = \xi_4(2,3) = 1 \quad . \quad (17)$$

Eq. (13) then requires

$$\xi_4(1,3)\xi_4(2,4) = 2 \quad . \quad (18)$$

There is an infinity of choices for $\xi_4(1,3)$ and $\xi_4(2,4)$. Whichever choice is made, imposes a specific relation among the corresponding free phases.

### 3. Quantized representation
### 3.1 The operator F

The quantized representation is obtained by constructing an operator F over a Fock space of particles (they may be either Bosons or Fermions) in terns of the number operators:

$$N_k = a_k^\dagger a_k \quad , \quad \begin{pmatrix} [a_k, a_{k'}^\dagger] = \delta_{k,k'} \\ \{a_k, a_{k'}^\dagger\} = \delta_{k,k'} \end{pmatrix} \quad . \quad (19)$$

In Eq. (19), $a_k$ and $a_k^\dagger$ are, respectively, particle annihilation and creation operators, [,] stands for the commutator of Bosonic operators, and {,} – for the anti-commutator of Fermionic operators.

Regarding the Fock space as a direct sum of subspaces with $N \geq 0$ particles, the operator F is diagonal in particle numbers, and given by:

$$F(t,x,y) = \begin{pmatrix} 1 & & & 0 \\ & F_1(t,x,y) & & \\ & & F_2(t,x,y) & \\ 0 & & & \ddots \end{pmatrix}. \qquad (20)$$

In Eq. (20), each operator, $F_N(t,x,y)$, acts on the subspace of $N \geq 1$ particles. $F_N(t,x,y)$ are given by:

$$F_1(t,x,y) = \int_{-\infty}^{\infty} dk_1 \, N_{k_1} \exp(\theta_1)$$
$$F_{N \geq 2}(t,x,y) = \frac{1}{(N!)} \int_{-\infty}^{\infty} dk_1 \int_{-\infty}^{\infty} dk_2 \ldots \int_{-\infty}^{\infty} dk_N \prod_{i=1}^{N} \{N_{k_i} \exp(\theta_i)\} \prod_{1 \leq j < l \leq N} |k_l - k_j| \qquad (21)$$

To circumvent the divergence of integrals in Eq. (21) as $|k| \to \infty$, one has to employ a regularization procedure. One may truncate the integrals at a high cut-off, $k = \pm K$, or multiply each exponential wave function by a factor, which falls off sufficiently rapidly as $|k| \to \infty$, e.g., $\exp(-\alpha k^4)$. When matrix elements of the operator F are calculated, the $\delta$ - functions in the commutators or anti-commutators of Eq. (19) eliminate all integrals and yield the integrands. In the cut-off procedure, the result does not depend on $K$, provided $K$ is sufficiently high. Using multiplicative damping factors, these amount to phase shifts in the resulting solitons [89]. One may then set $\alpha$ to zero. In the following, regularization is implied in all statements regarding operators.

**3.2 The wave function**
A state in the Fock space, in which a single wave number, $q$, is occupied by $n_q \geq 1$ particles (clearly, for Fermions, only $n_q = 1$ is allowed), will be denoted by:

$$|\{q, n_q\}\rangle = \frac{(a_q^\dagger)^{n_q}}{\sqrt{n_q!}} |0\rangle . \qquad (22)$$

The notation for a state of $N$ particles distributed among several wave numbers is self-evident:

$$|\{q_1, n_{q_1}\}, \ldots, \{q_P, n_{q_P}\}\rangle = \prod_{j=1}^{P} \frac{(a_{q_j}^\dagger)^{n_{q_j}}}{\sqrt{n_{q_j}!}} |0\rangle , \quad \left(\sum_{j=1}^{P} n_j = N\right) . \qquad (23)$$

In the case of Bosons, occupation of a wave number, $q$, by $n_q > 1$ bosons, amounts to a mere phase shift in the soliton solution of the classical evolution equation [89]. Therefore, in the following, attention will be focused on states, in which each wave number is occupied by a single particle.

The "pure" $N$-particle states are eigenstates of the operator F:

$$F|\{q_1,1\},....,\{q_N,N\}\rangle = \begin{cases} \exp(\theta_1)|\{q_1,1\}\rangle & N = 1 \\ \left(\prod_{i=1}^{N}\exp(\theta_i)\prod_{\substack{1 \le j \le N \\ j \ne i}}|q_i - q_j|\right)|\{q_1,1\},....,\{q_N,N\}\rangle & 2 \le N \le M-1 \end{cases}. \quad (24)$$

However, the eigenvalues do not coincide with the function, $f$, of Eq. (7), which yields soliton solutions. (This is an important difference between the present case and that of the equations in (1+1) dimensions, where the eigenvalues are equal to the Hirota function, $f(t,x)$ [89].)

The discussion in Section 2.4 indicates that the coefficients, $\xi_M(i_1,....,i_N)$, in Eq. (7) characterize a specific soliton solution. Hence, in the spirit of separation in Eq. (3) between the operator, U, which is equation specific, and the state in the Fock space, which is soliton-solution specific, the coefficients, $\xi_M(i_1,....,i_N)$, are incorporated in the construction of a "mixed" $N$-particle state:

$$|\psi_N(q_1,...,q_M)\rangle = \sum_{1 \le i_1 < .... < i_N \le M} \sqrt{\xi_M(i_1,....,i_N)} |\{q_{i_1},1\},....,\{q_{i_N},1\}\rangle . \quad (25)$$

$\xi_M(i_1,....,i_N)$ must obey Eq. (9). They may obtain any values for $(M,N) = (M,1)$ and $(M,M-1)$. For $\{M \ge 4, 2 \le N \le M-2\}$, they are constrained by bilinear relations, such as Eqs. (14) and (15). Finally, for the wave function of Eq. (25) to be properly normalized, one has to require

$$\sum_{1 \le i_1 < .... < i_N \le M} \xi_M(i_1,....,i_N) = 1 . \quad (26)$$

This requirement amounts to multiplying the function $f(t,x,y)$ of Eqs. (5) and (7) by a normalization constant, an operation that does not affect the solution, $u(t,x,y)$, of Eq. (4).

### 3.3 The classical solution
Calculation of the matrix element of the operator $F(t,x,y)$ of Eqs. (20) and (21) in the state $|\psi_N(q_1,...,q_M)\rangle$ of Eq. (25) yields the function $f_{\bar{\xi}}(M,N)$, corresponding to a solution of Eq. (1):

$$\langle\psi_N(q_1,...,q_M)|F(t,x,y)|\psi_N(q_1,...,q_M)\rangle = f_{\bar{\xi}}(M,N) \ . \tag{27}$$

### 3.4 The operator U
The operators $F(t,x,y)$ and $\partial_x F(t,x,y)$ commute, and $F(t,x,y)$ is positive definite. Hence, the operator version of Eq. (5) yields the operator $U(t,x,y)$:

$$U(t,x,y) = 2\partial_x\left(F(t,x,y)^{-1}\partial_x F(t,x,y)\right) \ . \tag{28}$$

Using Eqs. (27) and (28), the solution of Eq. (4) is obtained as the matrix element of $U(t,x,y)$ in $|\psi_N(q_1,...,q_M)\rangle$:

$$\langle\psi_N(q_1,...,q_M)|U(t,x,y)|\psi_N(q_1,...,q_M)\rangle = u_{\bar{\xi}}(M,N) \ . \tag{29}$$

Clearly, on the states $\psi_N(q_1,...,q_M)$, $U(t,x,y)$ obeys the operator version of Eq. (4):

$$\frac{\partial}{\partial x}\left(-4\frac{\partial U}{\partial t} + \frac{\partial^3 U}{\partial x^3} + 6U\frac{\partial U}{\partial x}\right) + 3\frac{\partial^2 U}{\partial y^2} = 0 \ . \tag{30}$$

The Hamiltonian density leading to Eq. (28) is the operator analog of its classical version [88]:

$$H[U] = \frac{1}{2}(U_x)^2 - U^3 + \frac{3}{2}U\partial_x^{-2}U_{yy} \ . \tag{31}$$

Finally, the matrix element of $H[U]$ in $|\psi_N(q_1,...,q_M)\rangle$ yields the classical Hamiltonian, $H[u]$:

$$\langle\psi_N(q_1,...,q_M)|H[U]|\psi_N(q_1,...,q_M)\rangle = H\left[u_{\bar{\xi}}(M,N)\right] \ , \quad H[u] = \frac{1}{2}(u_x)^2 - u^3 + \frac{3}{2}u\partial_x^{-2}u_{yy} \ . \tag{32}$$

## 4. Concluding comments

This paper has presented the construction of a quantized representation for the classical KP equation dimensions over the sector of soliton solutions. The separation in Eq. (3) between the operator U, which is equation-specific, and the $N$-particle state, $|\psi_N\rangle$, which is in one-to one correspondence with a soliton solution in the study of equations in (1+1) dimensions [89], is maintained here as well. This separation requires that the coefficient vector, $\vec{\xi}_M$, be incorporated in the definition of a "mixed" $N$-particle state, $|\psi_N(q_1,...,q_M)\rangle$, which yields the classical solution through Eq. (27), and *not* in the definition of the operator, $F(t,x,y)$. Here a difference between the (1+1)- and (2+1)-dimensional cases emerges. In the (1+1)-dimensional case, the exponential terms may be multiplied by arbitrary (positive) coefficients, which can be absorbed as free initial phase shifts, and the $N$-particle states are "pure" states of the form $|\{q_1,1\},...,\{q_N,1\}\rangle$. As pointed out in Section 2.4, the $\vec{\xi}_M$-coefficients cannot be absorbed in the exponential factors as *arbitrary* initial phase shifts, $\theta_i^0$. The bilinear constraints (e.g., Eqs. (14) or (15)) that the components of $\vec{\xi}_M$ must obey restrict some of the freedom in the initial phases.

Having constructed the operator $U(t,x,y)$, any property that the classical solution has may be now given an operator analog. Finally, the construction of the quantized representation presented here opens the door to the addition of perturbations that do not have classical analogs [89].

Finally, the ideas presented in [89] and in this paper raise the possibility that quantization of nonlinear evolution equations, which traditionally is achieved through the introduction of filed operators (linear in the creation and annihilation operators) may be very easily achieved through the introduction of the composite operator, U, which is constructed in terms of the number operator (bilinear in the creation and annihilation operators).

Acknowledgment: The author is thankful for critical reviews made by referees, and thanks G. Biondini and B. Prinari for directing his attention to the issue of the quantization of the KP equation.